## Solvent-Mediated Disintegration of a Vitrimer Network

Manoj K Patel and Tarak K Patra*

Department Of Chemical Engineering
Indian Institute of Technology Madras, Chennai, TN-600036, India

## Abstract

Vitrimers are dynamic polymer networks characterized by associative bond exchange, which enables topological rearrangement while preserving the overall network connectivity. However, how this dynamic connectivity evolves in the presence of a solvent, particularly as a function of temperature, remains poorly understood. Here, we employ hybrid molecular dynamics–Monte Carlo (MD-MC) simulations to elucidate the temperature-dependent restructuring and dissociation of vitrimer networks in solution. By constructing temperature–composition phase diagrams, we demonstrate that the critical temperature for network dissolution increases nonlinearly with crosslink concentration. We find that the network disintegration begins well below the critical dissolution temperature, indicating that the loss of network connectivity can occur even before the onset of macroscopic dissolution. However, the network undergoes extensive restructuring around the critical temperature for a highly crosslinked system, resulting in the formation of numerous smaller, disconnected networks. Concurrently, the number of intrachain dynamic bonds increases with temperature. Interestingly, complete disintegration into individual polymer chains is not observed; instead, the solution comprises a broad distribution of smaller network fragments. These findings reveal the complex interplay between solvent-induced network disintegration and dynamic bond redistribution, providing molecular-level insights into the dissolution of vitrimer networks and their potential for solvent-assisted recycling.

**Keywords:** Bond exchange reaction, Vitrimers, Network disintegration, Swelling, and Chemical recycling

*Author to correspond, E-mail: tpatra@iitm.ac.in

Vitrimers are polymer networks that are formed through reversible covalent bonds between individual polymers. Unlike conventional thermosets with permanently fixed network structures, vitrimers undergo bond-exchange reactions that allow network topology to rearrange without substantially altering the overall connectivity. This dynamic rearrangement of crosslink bonds provides a unique set of properties, including reprocessability, self-healing, weldability, adhesion, and shape memory.[1–11] Dynamic bond exchange is mediated by various chemical reactions, including transesterification, disulfide exchange, and imine exchange.[12,13] The persistence of network connectivity in vitrimers has traditionally led to the notion that these materials remain intrinsically insoluble, even under conditions that induce substantial swelling.[14] However, an increasing body of experimental evidence contradicts this conventional view, demonstrating that vitrimer networks can undergo partial, and under certain conditions, extensive loss of connectivity.[15–18] For instance, an amine–epoxy thermoset incorporating dynamic disulfide bonds has been reported to undergo complete degradation in a thiol-containing solvent (2-mercaptoethanol) at room temperature.[19] Similarly, selective solubility has been demonstrated in an imine-based network in tetrahydrofuran,[20] while an imine-containing epoxy system has been shown to dissolve completely in dichloromethane, offering a potential route for solvent-assisted recycling.[21] Vanillin-based polyimine vitrimers have likewise been successfully recycled using suitable solvents,[22] while studies on polybutadiene-based vitrimers have revealed a substantial population of topological defects that may compromise network integrity.[23] Taken together, these findings point to a more complex interplay between network swelling, topological defects, and dissolution**.** In particular, they raise fundamental questions about how a vitrimer network evolves upon exposure to a solvent: does swelling precede the loss of network connectivity? And under what conditions does a swollen network transition into a partially or completely dissolved state? Understanding the interplay between these processes is essential for establishing the molecular mechanisms governing vitrimer dissolution and, ultimately, developing strategies for their solvent-assisted recycling**.**

Prior theoretical and simulation studies by Smallenburg, Leibler, and Sciortino, using a patchy-particle model, demonstrated the coexistence of a macroscopic network phase with a dilute phase containing finite-sized networks.[24] Computer simulations by Kumar and Panagiotopoulos further established that physically associating polymers undergo a sol–gel transition.[25] In addition, the equilibrium mean-field theory of Semenov and Rubinstein predicts an intrinsic thermodynamic tendency toward phase separation associated with the formation of reversible networks through molecular association.[26] Building on these insights, we investigate how the connectivity and structural integrity of vitrimer networks evolve with temperature in an explicit solvent environment. Specifically, we address several fundamental questions: Does a vitrimer network undergo a sharp transition from a percolated to a disconnected state, or does its connectivity progressively deteriorate over a broad temperature range? Does network disintegration precede the apparent critical conditions for macroscopic dissolution, and, if so, how does the onset of connectivity loss relate to the dissolution temperature? Furthermore, how are crosslinks redistributed among individual polymer chains during dissolution, and how does this redistribution influence the progressive disintegration of the network? Addressing these questions is essential for establishing a more comprehensive molecular-level understanding of the network disintegration in a thermochemical environment. These questions are also important from a materials-design perspective. Chemical durability—the ability of a vitrimer network to resist unintended bond cleavage and loss of connectivity—is central to its reliability and long-term performance in demanding applications, including structural composites, coatings, adhesives, and biomedical materials. At the same time, network rearrangement and disintegration are essential for reprocessing and chemical recycling. Thus, the dynamic covalent chemistry that gives vitrimers their recyclability may also introduce pathways for network degradation under unfavorable thermal or chemical conditions. Understanding this trade-off between durability and disintegration of a network is therefore critical for designing vitrimer networks with predictable service lifetimes and controllable end-of-life behavior.

Here, we address these questions using a hybrid MD–MC simulation framework to investigate the temperature-dependent structural evolution and dissolution of vitrimer networks in an explicit solvent environment. By systematically varying the temperature, we construct solution-phase diagrams for a given crosslink concentration, delineating the regimes of network integrity, partial disintegration, and dissolution. We identify a critical temperature associated with network dissolution and demonstrate that its dependence on crosslink concentration is distinctly nonlinear, highlighting the complex interplay between network connectivity and polymer–solvent interactions. Interestingly, we find that the loss of network connectivity sets in at temperatures substantially below the critical dissolution temperature, indicating that the disintegration of the network is a progressive process rather than an abrupt transition. However, as the temperature approaches the critical value, network fragmentation becomes considerably more pronounced, accompanied by the formation of smaller-sized nanostructures. Thus, even above the critical temperature, the system does not necessarily comprise completely isolated polymer chains; instead, it produces smaller network fragments. We further examine the temperature-dependent redistribution of dynamic crosslinks among polymer chains and find that the average number of dynamic bonds per chain increases with temperature. This behavior suggests an increasing contribution from intrachain associations relative to interchain crosslinks, consistent with the progressive fragmentation of the extended network into smaller connected structures. Overall, these findings establish a molecular-level picture of vitrimer dissolution in which network disintegration, crosslink redistribution, and the emergence of finite-sized polymer clusters (nanoparticles) occur over a broad temperature range. This provides new insights into the temperature-dependent chemical stability and reprocessability of vitrimer networks and highlights the importance of considering network connectivity beyond the conventional macroscopic criterion for dissolution.

We perform molecular dynamics (MD) simulations in an isothermal-isobaric ensemble. The initial simulation setup consists of a polymer slab sandwiched between two blocks of solvent particles along the $z$-axis, as shown in Figure 1. A polymer is modelled as the Kremer-Grest bead-spring chain.[27] In this model, all non-bonded monomers interact via the Lennard-Jones (LJ) potential of the form $V(r) = 4\epsilon\left[\left(\frac{\sigma}{r}\right)^{12} - \left(\frac{\sigma}{r}\right)^{6}\right]$, with a cutoff distance of $r_c = 2.5\sigma$, where $\sigma$ and $\epsilon$ represent the monomer size and interaction energy scale, respectively. A pair of bonded monomers in a polymer chain interacts via the finite extensible nonlinear elastic (FENE) potential, $V_{bond}(r_{ij}) = -0.5kR_0^2 \ln\left[1 - \left(\frac{r_{ij}}{R_o}\right)^2\right] + 4\epsilon\left[\left(\frac{\sigma}{r}\right)^{12} - \left(\frac{\sigma}{r}\right)^{6}\right] + \epsilon$, with a maximum bond extension of $R_0 = 1.5\sigma$ and a spring constant of $k = 30\epsilon/\sigma^2$. The 2nd term in the $V_{bond}(r_{ij})$ is truncated at a distance $r = 2^{1/6}\sigma$. A solvent particle is also modelled as an LJ particle with identical size and energy parameters as that of a polymer monomer. The monomer-solvent interaction is also modelled via the LJ interaction with the interaction energy $\epsilon_{ps} = 0.9\epsilon$. The system consists of 19190 particles, including solvent particles and 213 polymer chains. Each polymer chain comprises 30 monomers. The simulation box is periodic in all three directions. All systems are equilibrated for $10^7$ MD steps with an integration timestep of $\Delta t = 0.005\tau$. Here, $\tau = \sqrt{m\sigma^2/\epsilon}$ is the unit of time, with $m$, $\sigma$ and $\epsilon$ are the mass, size and interaction energy of a monomer, respectively. The reduced temperature and pressure of the system are $\frac{k_B T}{\epsilon} = 1$ and $\frac{P\sigma^3}{\epsilon} = 1$, respectively. Here, $T$, $P$ and $k_B$ are the actual temperature, pressure and the Boltzmann constant, respectively. The pressure and temperature are maintained by the Nosé-Hoover barostat and thermostat, respectively. The pressure is controlled only via the z-axis. In the subsequent sections, the reduced temperature is abbreviated as $T$.

After this initial MD run, a target number of crosslink bonds between monomers is randomly generated, and we perform a hybrid MC-MD simulation to generate a vitrimer trajectory. The MC-MD method is adopted from the work of Li *et al.*[28] and has been implemented for various associative network systems in our previous works.[29–31] The monomers that participate in crosslink bond formation are labelled as stickers. The crosslink bond between two stickers is also modelled as the FENE bond. In this hybrid approach, we perform 100 MC moves (one MC cycle) at regular intervals of the MD simulation. The

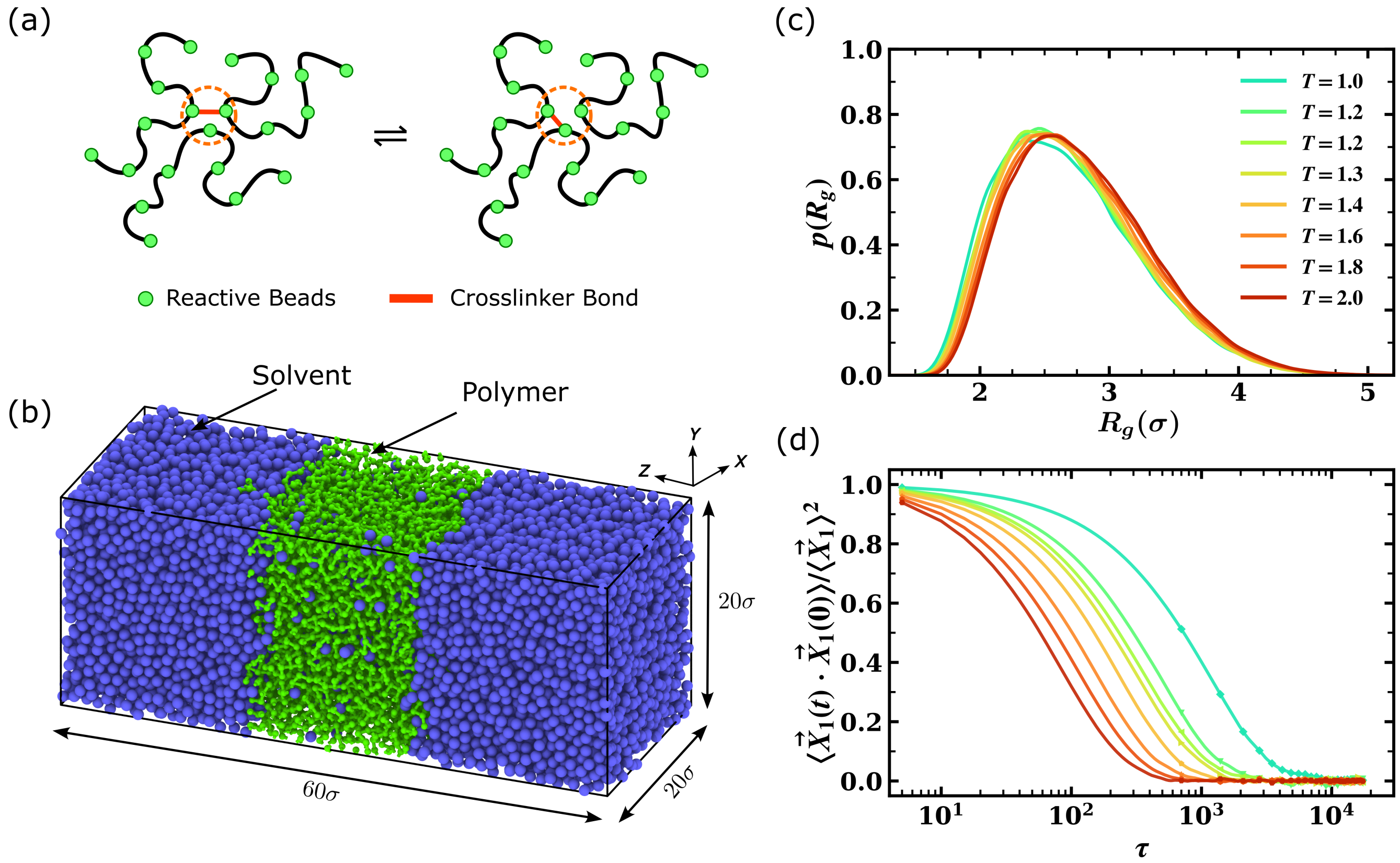

*Figure 1: (a) Schematic representation of the dynamic crosslink exchange in a vitrimer network, showing the bond swap/exchange between reactive beads. (b) Initial snapshot of the simulation box featuring the polymer slab ($C_{xl}$ = 0.03) exposed to solvent along the z-axis. (c) Probability distribution of the radius of gyration, $p(R_g)$ for the system, evaluated across reduced temperatures from $T$ = 0.8 to 1.7. (d) Normalized time autocorrelation functions of the longest Rouse mode of the system at the corresponding temperatures.*

MD time elapsed between two consecutive MC cycles is denoted as $\tau_c$, and is set to $5\tau$. At each MC step, a sticker particle selects a new bonding partner and exchanges its existing bond with the new one according to the Metropolis acceptance criterion, as schematically shown in Figure 1a. Here, each sticker can participate in at most one crosslink. We note that all monomers are eligible to participate in crosslink formation. The crosslink density $C_{\text{xl}}$ is defined as the ratio of the number of crosslinked monomers to the total number of monomers. The crosslink density $C_{\text{xl}}$ is kept constant throughout a simulation. We perform simulations across five distinct crosslink densities: $C_{\text{xl}} = 0.00$ (no crosslink), 0.03 (213 crosslinks), 0.12 (771 crosslinks), 0.16 (1080 crosslinks) and 0.21 (1374 crosslinks). Each system is studied over a wide range of temperatures. A simulation is carried out for $10^7$ MD equilibration steps, followed by another $10^7$ steps of production run. All MD simulations are conducted using the LAMMPS[20] open-source code. The polymer chain radius of gyration ($R_g$) distribution across temperatures for a representative crosslink density, $C_{\text{xl}} = 0.03$ is shown in Figure 1c. The average radius of gyration ($R_g$) is about $2.5\sigma$, and the lowest simulation box dimension is $L_x = L_y = 20\sigma$. As $R_g$ is substantially smaller than the box length, we do not expect any significant system size effect in our analysis. The autocorrelation function of the longest Rouse mode of the system is shown in Figure 1d for a representative case of $C_{\mathrm{xl}} = 0.03$. The autocorrelation functions decay fully within our simulation time. This confirms that the chains are fully relaxed and that the simulation times are sufficient to achieve equilibration.

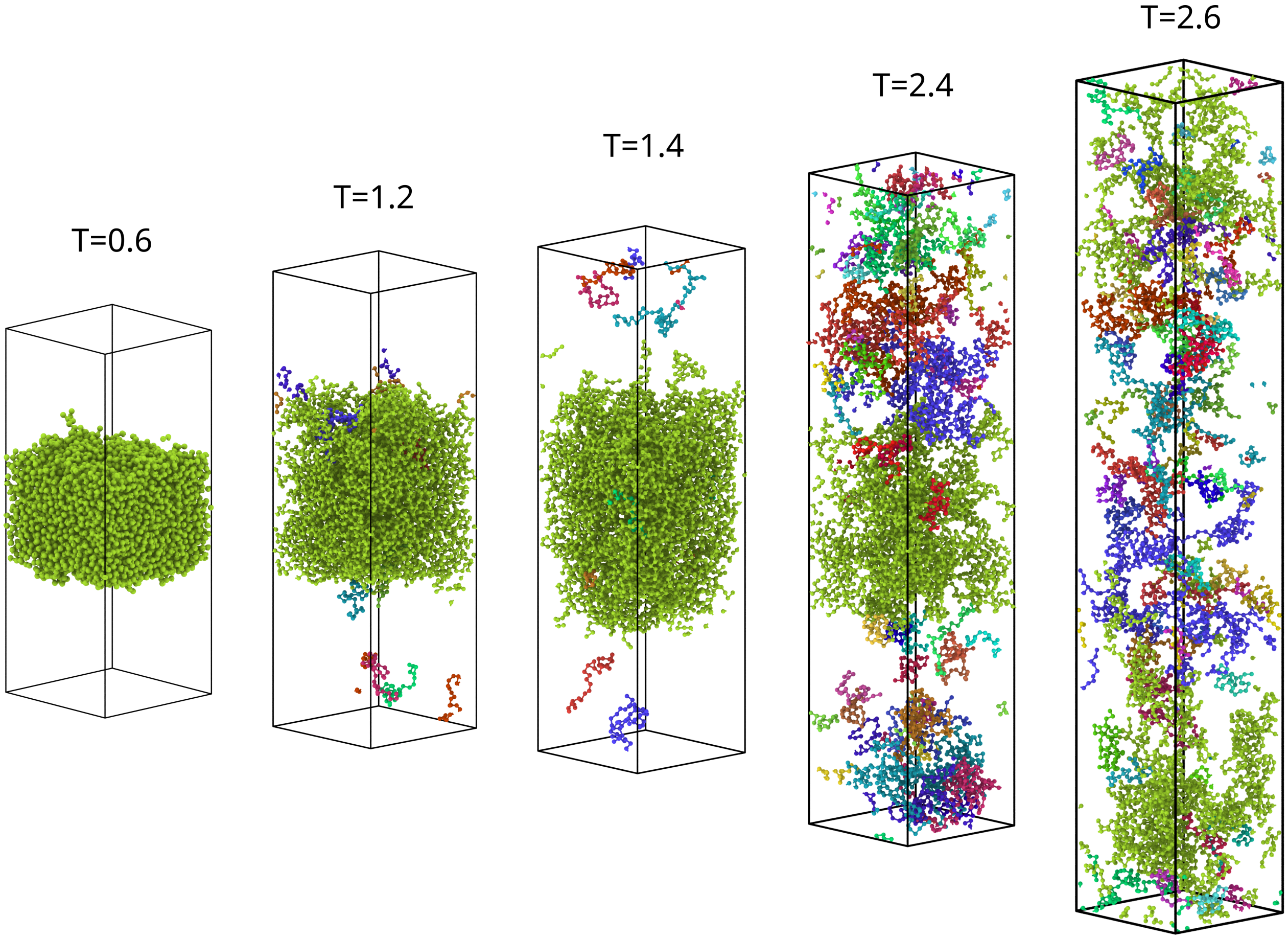

*Figure 2 Representative simulation snapshots illustrating the equilibrated state and phase behavior of the polymer vitrimer solution at a crosslink density of $C_{xl} = 0.12$ across temperatures from $T = 0.8$ to 2.6. For visualisation and clarity purposes, solvent molecules have been removed from the snapshots.*

Figure 2 displays equilibrated MD snapshots of the system for a representative crosslink concentration of $C_{\mathrm{xl}} = 0.12$ across various reduced temperatures ranging from $T = 0.8$ to 2.6. For low temperatures ($T = 0.8$), the polymer chains form a highly connected network. As the temperature increases, numerous chains are found to detach from the primary network. To visually track network disintegration, a cluster-based color-coding scheme is employed: all chains connected by crosslinks share a uniform color. Consequently, the proliferation of varied colors at elevated temperatures provides a clear visual indicator of progressive breakdown of a network into multiple sub-networks. Chains from the polymer

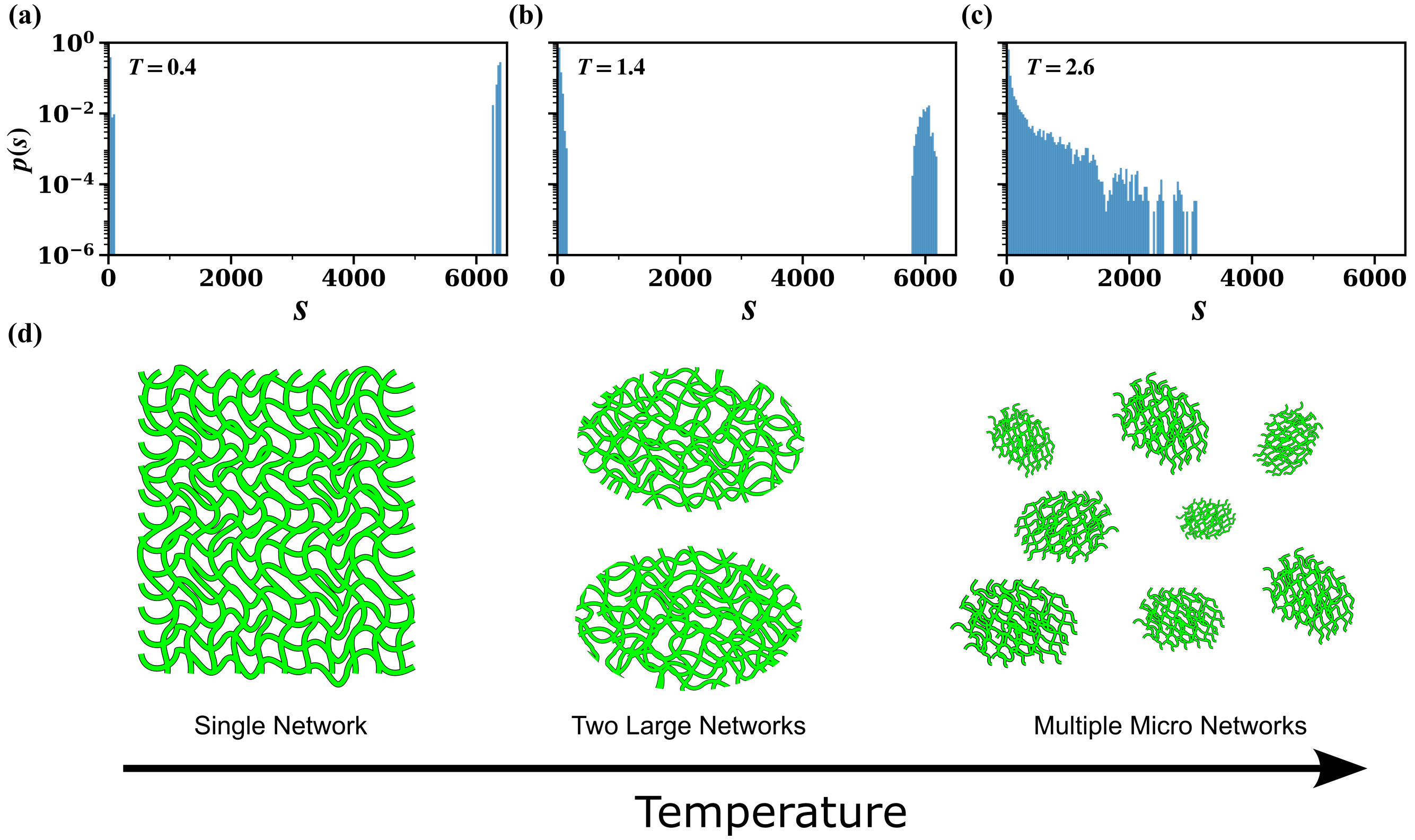

*Figure 3: Network distributions are shown in (a), (b) and (c) for three different temperatures for a crosslink concentration $C_{xl}$= 0.12. Panel (d) is a schematic representation of a network that fragments into smaller networks as the temperature increases.*

reach- domain migrate to the solvent phase, either remaining as single chains or connecting with other free chains to form networks. The size of these networks varies, and their number increases with temperature. We quantify this disintegration process by monitoring the size distribution of these networks. The size of a network (*s)* is defined as the number of monomers in it. Any two monomers that are connected by a FENE bond (inherent static bond or dynamic crosslink bond) are part of a network. The network size distribution is shown in Figure 3 for the crosslink concentration $C_{\mathrm{xl}} = 0.12$ for a few selected temperatures. At very low temperatures, it is predominantly a single large network in which most chains are connected by dynamic crosslinks. The network distribution exhibits two sharp peaks at the extreme values of the network size for $T = 0.4$. The peak at low values of $s$ is due to the transient or occasional detachment of a few chains from the large network at the network–solvent interfacial region. As the temperature increases, the primary network fragments and produces multiple networks. The cluster size distribution shifts to the left and becomes very broad at high temperatures, as shown for T=2.6 in Figure 3c. The fragmentation is schematically depicted in Figure 3d.

We further estimate the monomer densities in the polymer-rich phase and solvent-rich phase and construct a phase diagram (cf. Figure 4a). As shown in the inset of Figure 4a, the critical temperature $T_c$ increases rapidly with increasing $C_{\mathrm{xl}}$, demonstrating that denser crosslinked networks require greater thermal energy to dissolve in the solution. We monitor the fraction of chains contained within the largest spanning network ($f_{NET}$) as a function of the temperature (Figure 4b). For highly crosslinked systems ($C_{\mathrm{xl}} = 0.12, 0.17$ and $0.21$), at low temperatures, all the 213 chains remain connected within a single network ($f_{\mathrm{NET}} = 1.0$). As the temperature increases, the primary network swells, thereby reducing the local density of monomers. The primary network dissociates by 20% until Tc. Subsequently, it undergoes a rapid disintegration at and above Tc. This is consistent with recent experiments suggesting that dissolution is significantly faster near the transition point.[32] We have also considered a case where the crosslink concentration is very low, leading to a distribution of networks within the polymer phase. In this low crosslink concentration ($C_{\text{xl}} = 0.03$), the system exhibits $f_{\text{NET}} \approx 0.4$ at T=0.5, indicating that roughly 40% of chains are part of the largest network. Near the critical temperature $T_c$, the largest network consists of about 3% of the system's chains. Therefore, a significant dissociation occurs below the $T_c$, and there is no significant change in the network distribution above the $T_c$. for this low crosslink limit. We quantify the presence of single chains as a function of temperature (Figure 4c) for all the cases. The number of free chains in the system grows rapidly for low crosslink density. In the case of highly crosslinked systems, it grows slowly with temperature. Around 10% of chains are free at $T_c$ for the highly crosslinked systems. It becomes about 30% at $T/T_C = 2.0$. This disintegration behavior is further corroborated by the average number of networks in the system, as shown in Figure 4d. The number of networks increases monotonically with temperature, suggesting a larger network breaks into a larger number of smaller-sized networks.

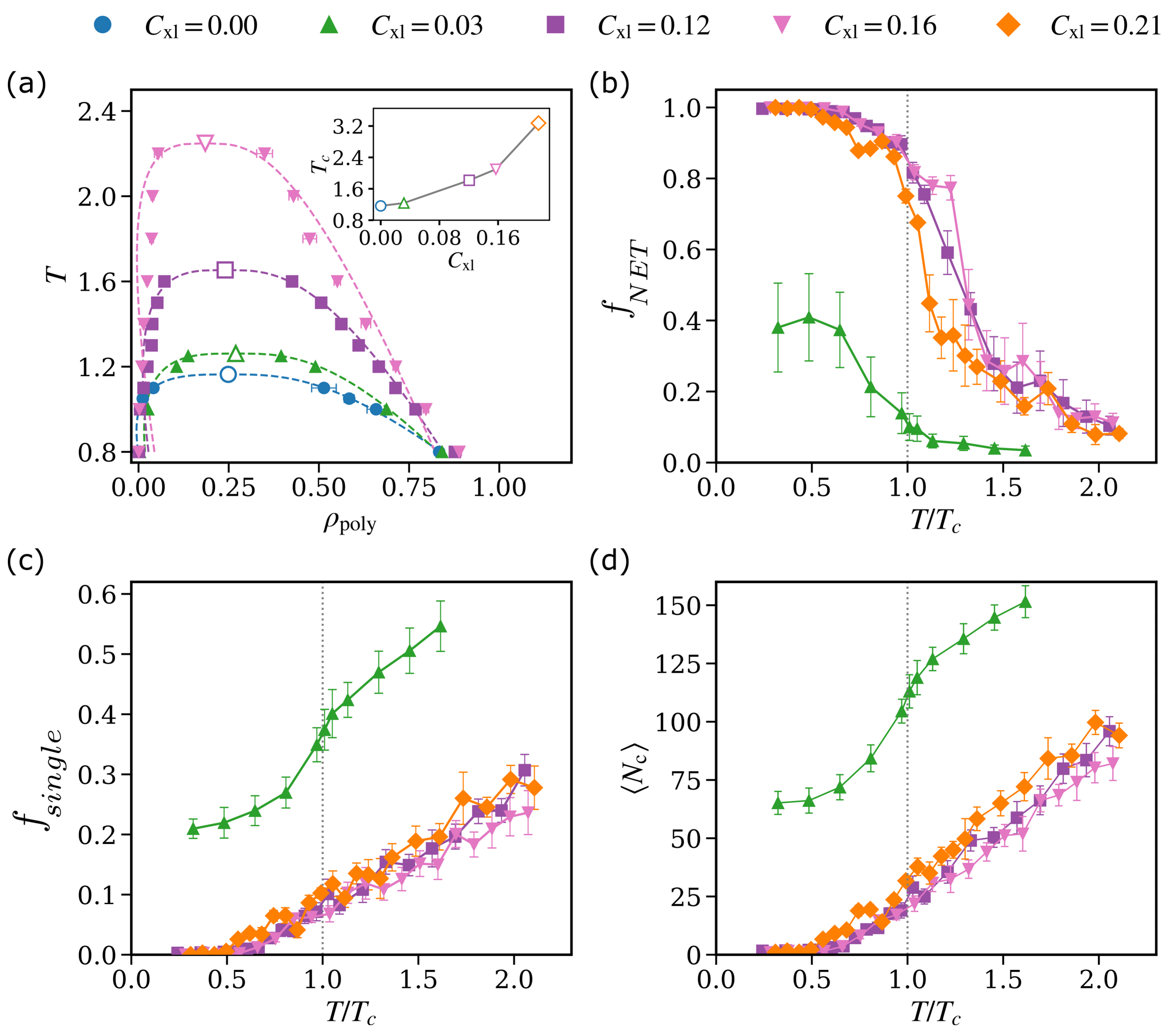

*Figure 4: Network dissolution and dissociation. (a) Polymer phase diagram. The inset represents the critical temperature as a function of the conrsslink concentration in the system. (b) Fraction of freely floating single chains ($f_{single}$) as a function of normalized temperature ($T/T_c$). (c) Fraction of polymer chains contained within the largest spanning network ($f_{NET}$) is plotted as a function of $T/T_c$. (d) Mean network size as a function of normalized temperature ($T/T_c$) across different $C_{xl}$ cases.*

Moreover, we examine the fractions of inter-chain, $< C_{\text{xl-inter}} >$ and intra-chain, $< C_{\text{xl-inter}} >$ crosslink bonds as a function of normalized temperature ($T/T_c$) (cf. Figure 5). Our simulations maintain a fixed total crosslink density ($C_{\text{xl}}$), the redistribution between intermolecular and intramolecular bonding provides direct insight into the structural evolution of the system. At low temperatures ($T/T_c < 1$), where the polymer system exists majorly as a single, highly connected macroscopic network, the vast majority of crosslinks are inter-chain ($\sim 70-80\%$, Figure 5a), connecting distinct polymer chains across the percolated structure. As the temperature increases, the number of inter-chain crosslinks monotonically decreases. The fraction of inter-chain crosslinks decreases significantly down to $20-30\%$ (Figure 5a) at a very high temperature $T/T_c \sim 1.5$. As the crosslink density must remain constant, the detached chains and small oligomeric fragments form an increasing proportion of intra-chain crosslinks (changing from $\sim 20\%$ to nearly $80\%$ in Figure 5b). This pronounced crossover from inter-chain to intra-chain connectivity underscores how microscopic loop formation compensates for macroscopic network dissociation at elevated temperatures.

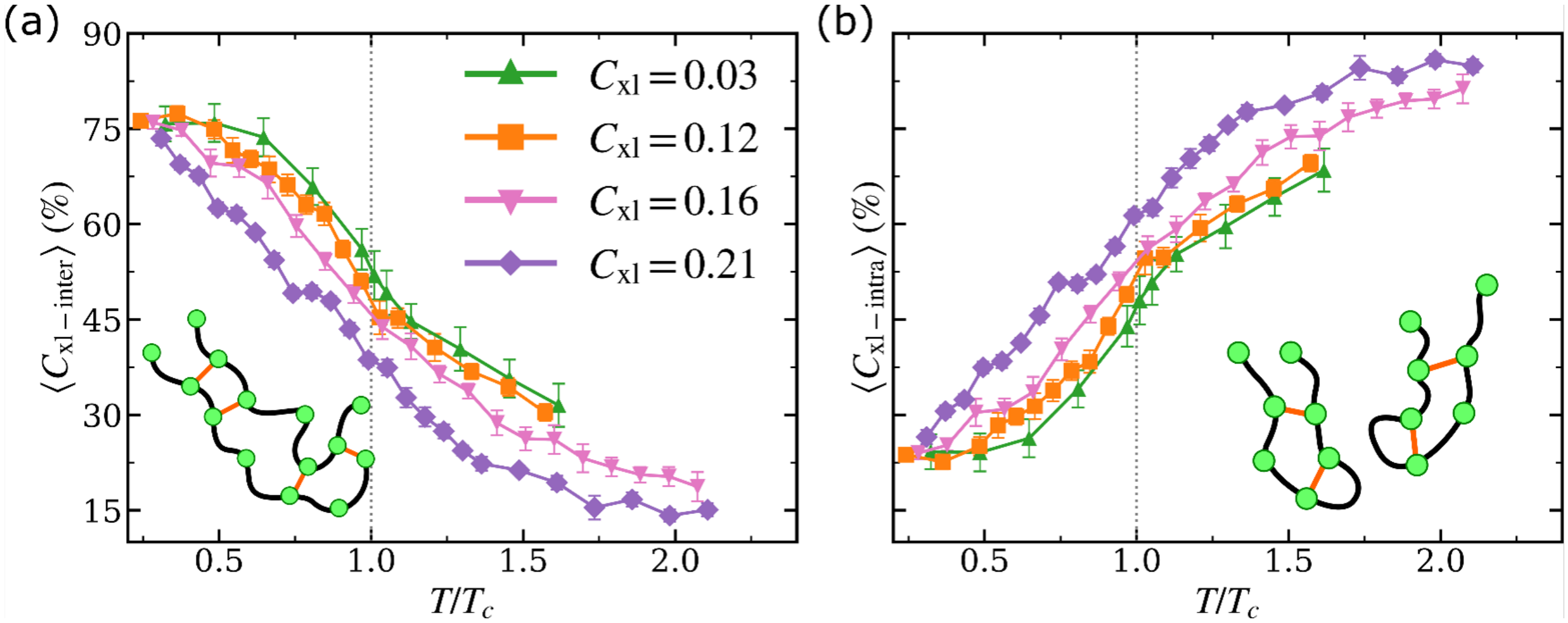

*Figure 5 The percentage of inter-chain crosslink bonds (a) and intra-chain crosslinks bonds (b) are plotted as a function of normalized temperature ($T/T_c$) for various crosslink densities ($C_{xl}$). The insets show schematics of two polymer chains containing crosslink bonds.*

In summary, we investigate the temperature-dependent dissolution and disintegration of vitrimer networks in an explicit solvent environment using a hybrid MC–MD simulation framework. We focus on how crosslink concentration and polymer–solvent interactions govern network swelling, structural integrity, and the progressive loss of network connectivity. We find that highly crosslinked vitrimer networks undergo substantial swelling while largely retaining their connectivity, with significant network disintegration occurring in the vicinity of the critical dissolution temperature. In contrast, networks with lower crosslink concentrations exhibit a more pronounced loss of connectivity as the temperature approaches the dissolution transition. The critical dissolution temperature increases monotonically and nonlinearly with crosslink concentration, highlighting the stabilizing effect of increased network connectivity against thermal disintegration. Importantly, network disintegration is initiated well below the critical dissolution temperature, demonstrating that the loss of macroscopic connectivity is a progressive process rather than an abrupt transition. With increasing temperature, the number of interchain crosslinks decreases approximately linearly, resulting in the fragmentation of the system-spanning network into smaller, dynamically connected subnetworks. Despite the constant overall crosslink density, the network undergoes substantial topological reorganization through the redistribution of dynamic bonds: interchain crosslinks are progressively converted into intrachain associations, leading to the formation of loops within individual polymer chains. This redistribution provides a molecular-level mechanism for the loss of network connectivity while preserving a finite population of dynamic bonds. These findings establish a molecular picture of vitrimer dissolution in which swelling, crosslink redistribution, and network fragmentation precede macroscopic dissolution. These insights advance the understanding of the temperature-dependent evolution of vitrimer network topology and the molecular mechanisms governing their stability, disassembly, and potential reprocessability in solution.

**Acknowledgement**

This work is made possible by financial support from the ANRF through an advanced research grant (ARG/2025/006228).